\documentclass[%
 reprint,
superscriptaddress,
 amsmath,amssymb,
pra,
floatfix,
]{revtex4-2}

\usepackage{graphicx}
\usepackage{dcolumn}
\usepackage{bm}

\usepackage{lipsum}  
\usepackage{color}

\newcommand{\centered}[1]{\begin{tabular}{l} #1 \end{tabular}} 

\begin{document}


\title{Solving two-dimensional quantum eigenvalue problems\\ using physics-informed machine learning}

\author{Elliott G. Holliday*}
\affiliation{Nonlinear Artificial Intelligence Laboratory, Physics Department, North Carolina State University, Raleigh, NC 27607, USA }

\author{John F. Lindner}
\affiliation{Nonlinear Artificial Intelligence Laboratory, Physics Department, North Carolina State University, Raleigh, NC 27607, USA }
\affiliation{Physics Department, The College of Wooster, Wooster, OH 44691, USA}

\author{William L. Ditto}
\affiliation{Nonlinear Artificial Intelligence Laboratory, Physics Department, North Carolina State University, Raleigh, NC 27607, USA }

\date{\today}

\begin{abstract}
A particle confined to an impassable box is a paradigmatic and exactly solvable one-dimensional quantum system modeled by an infinite square well potential. Here we explore some of its infinitely many generalizations to two dimensions, including particles confined to rectangle, elliptic, triangle, and cardioid-shaped boxes, using physics-informed neural networks. In particular, we generalize an unsupervised learning algorithm to find the particles' eigenvalues and eigenfunctions. During training, the neural network adjusts its weights and biases, one of which is the energy eigenvalue, so its output approximately solves the Schr\"odinger equation with normalized and mutually orthogonal eigenfunctions. The same procedure solves the Helmholtz equation for the harmonics and vibration modes of waves on drumheads or transverse magnetic modes of electromagnetic cavities. Related applications include dynamical billiards, quantum chaos, and Laplacian spectra.
\end{abstract}

\maketitle


\section{\label{sec:level1} Introduction}
Artificial intelligence has impacted our culture and livelihood dramatically since the turn of the millennium, from cellphones to self-driving cars and beyond.  Scientists have recently begun using machine learning techniques to not only improve our understanding of the world around us but change the way we approach scientific and computational methods, including those methods used to solve the fundamental differential equations that model physical phenomena in our world.

In previous work, physics-informed neural networks have been used to solve classical physics problems, with the Lagrangian and Hamiltonian formalisms, for both ordered and chaotic dynamics~\cite{Choudhary2020, Finzi2020A, Mattheakis2020B}. On a more fundamental level, methods have been developed to search for symmetries~\cite{Bondesan2019A}, conservation laws~\cite{Liu2021}, and invariants within such dynamical systems~\cite{Wetzel2020A}.  Furthermore, studies have been conducted on specific problems such as heat transfer~\cite{Cai2021}, irreversible processes~\cite{Lee2021A}, and energy-dissipating systems~\cite{Zhong2020A}. These techniques have even been applied to quantum systems~\cite{Cao2015,Raissi2019,Jin2020C,Jin2022C}. In this study, we use physics-informed neural networks to solve the quantum eigenvalue problem for particles confined to impassable planar boxes of diverse shapes.

Our work is an extension of that of Jin, Mattheakis, and Protopapas~\cite{Jin2020C,Jin2022C}, who use neural networks to solve the one-dimensional quantum eigenvalue problem for a small number of systems.  Here we extend the JMP algorithm to two-dimensions and find the eigenvalues and eigenfunctions of the Schr\"{o}dinger differential equation with Dirichlet boundary conditions in two-dimensional regions that exhibit classically regular or chaotic billiard dynamics, including the rectangle and cardioid.  

One of the most important features of JMP is the use of \textit{unsupervised learning}, so the neural network is not presented with the  solutions to the differential equation during training.  This characteristic showcases the natural ability of neural networks to find solutions to problems that may not be solvable analytically or numerically by other methods.  Extending this work to multi-dimensional systems advances both machine learning and physics by broadening the usefulness of neural networks and increasing the ways scientists can solve problems.

\section{Conventional Neural Networks}

\begin{figure}[b!]
    \centering
    \includegraphics[width=0.8\linewidth]{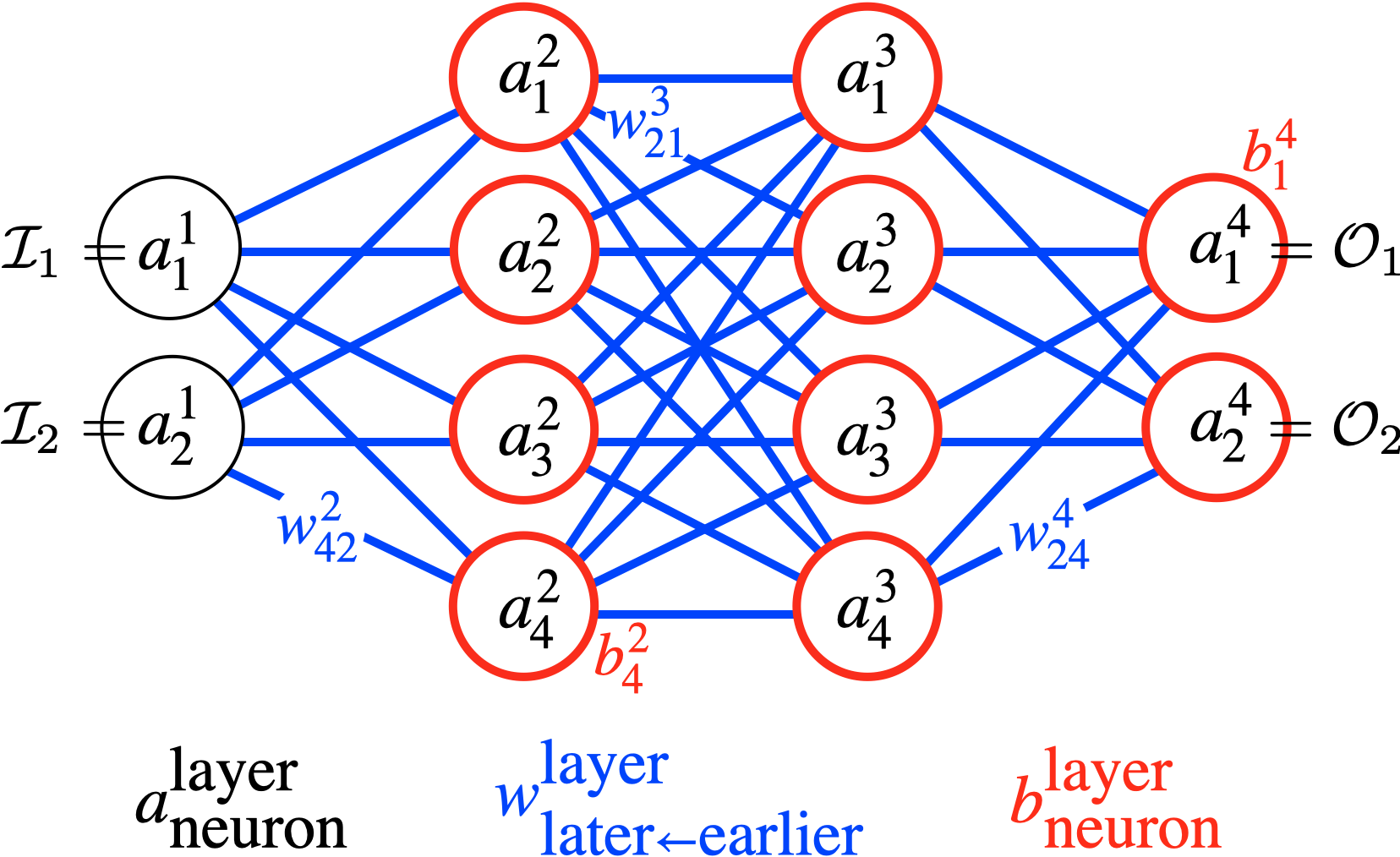}
    \caption{Conventional feed-forward 2:4:4:2 neural network.}
    \label{NN2442}
\end{figure}

Inspired by mammalian brains, conventional feed-forward neural networks are nested nonlinear functions that depend on many parameters called weights $w^l_{nm}$ and biases $b^l_n$, as in the Fig.~\ref{NN2442} schematic. Training adjusts the weights and biases to approximate desired outcomes.

Given a nonlinear neuron activation function like $\sigma[z] = \tanh z$, training recursively updates the neuron activities $ a^l_n = \sigma[z^l_n]$ according to their neuronal inputs
\begin{equation}
    z^l_n 
    = \sum_{m=1}^{N^{l-1}} w^l_{nm} \sigma[z^{l-1}_m] + b^l_n,
    \end{equation}
for neuron $1\le n\le N^l$ of layer $2\le l\le L$, where the network inputs and outputs are $\mathcal{I}_m = a^1_m$ and $\mathcal{O}_m = a^L_m$. As described in Appendix~\ref{GradientDescent}, stochastic gradient descent adjusts the weights and biases to minimize an error (objective, cost, loss) function like
\begin{equation}
    L = \sum_{s=1}^{N^{L}} \frac{1}{2} \left( \mathcal{O}_s - \hat{\mathcal{O}}_s \right)^2,
\end{equation}
where $\hat{\mathcal{O}}_s$ are the desired outputs.

\section{Methodology}

\subsection{1D Review}
We first review the formalism of Jin et al.~\cite{Jin2020C,Jin2022C} for the one-dimensional Schr\"odinger-equation eigenvalue-eigenfunction problem. Inputs of the neural network are positions $x$ and the constant 1, where the constant is converted to the energy eigenvalue $\lambda$ by an affine transformation. If the output of the neural network is the function $f(x,\lambda)$, then the eigenfunction
\begin{equation}
    \varPsi(x) = \varPsi_{b} + g(x)f(x,\lambda),
\end{equation}
where $\varPsi_{b}$ is the value of the function at the boundary, and $g(x) = 0$ on the boundary.

The loss function for this neural network incorporates the time-independent one-dimensional Schr\"{o}dinger equation,
\begin{equation} \label{TISE}
    H\varPsi(x) 
    = \left(-\frac{\partial^2}{\partial x^2} + V(x)\right)\varPsi(x) 
    = E\varPsi(x),
\end{equation}
where $V(x)$ is the potential energy function, and $\varPsi(x)$ can be taken to be real-valued. Once $\varPsi(x)$ is constructed, the appropriate derivatives are calculated using automatic differentiation. The loss function to be minimized is
\begin{equation}
    L = \langle\left(H\varPsi - \lambda\varPsi\right)^2\rangle_{x} + L_{\text{reg}},
\end{equation}
where $\langle\cdot\rangle_{x}$ means averaging over position $x$, and the regularization $L_{\text{reg}}$ has two components, $L_{\text{norm}}$ and $L_{\text{ortho}}$, which facilitate the search for non-trivial higher eigenvalues and eigenfunctions using physics-guided intuition.

The first part of the regularization is based on the normalization property of quantum eigenfunctions, $\varPsi \cdot \varPsi < \infty$. This is enforced by the loss function
\begin{equation}
    L_{\text{norm}} = \left(\varPsi \cdot \varPsi - \frac{M}{\Delta l}\right)^2,
\end{equation}
where $M$ is a normalization constant, and $\Delta l$ is the potential function length scale.  This portion of the loss function discourages the neural network from finding the trivial identically-zero solution. The second part of the regularization is based on the orthogonality property of quantum eigenfunctions, $\varPsi_1 \cdot \varPsi_2 = 0$. This is enforced by the loss function
\begin{equation}
    L_{\text{ortho}} = \varPsi_{s}\cdot\varPsi,
\end{equation}
where $\varPsi_{s}$ is the sum of the previously learned eigenfunction and $\varPsi$ is the eigenfunction being currently learned.  


\subsection{2D Extension}

\begin{figure}[b]
    \centering
    \includegraphics[width=1.0\linewidth]{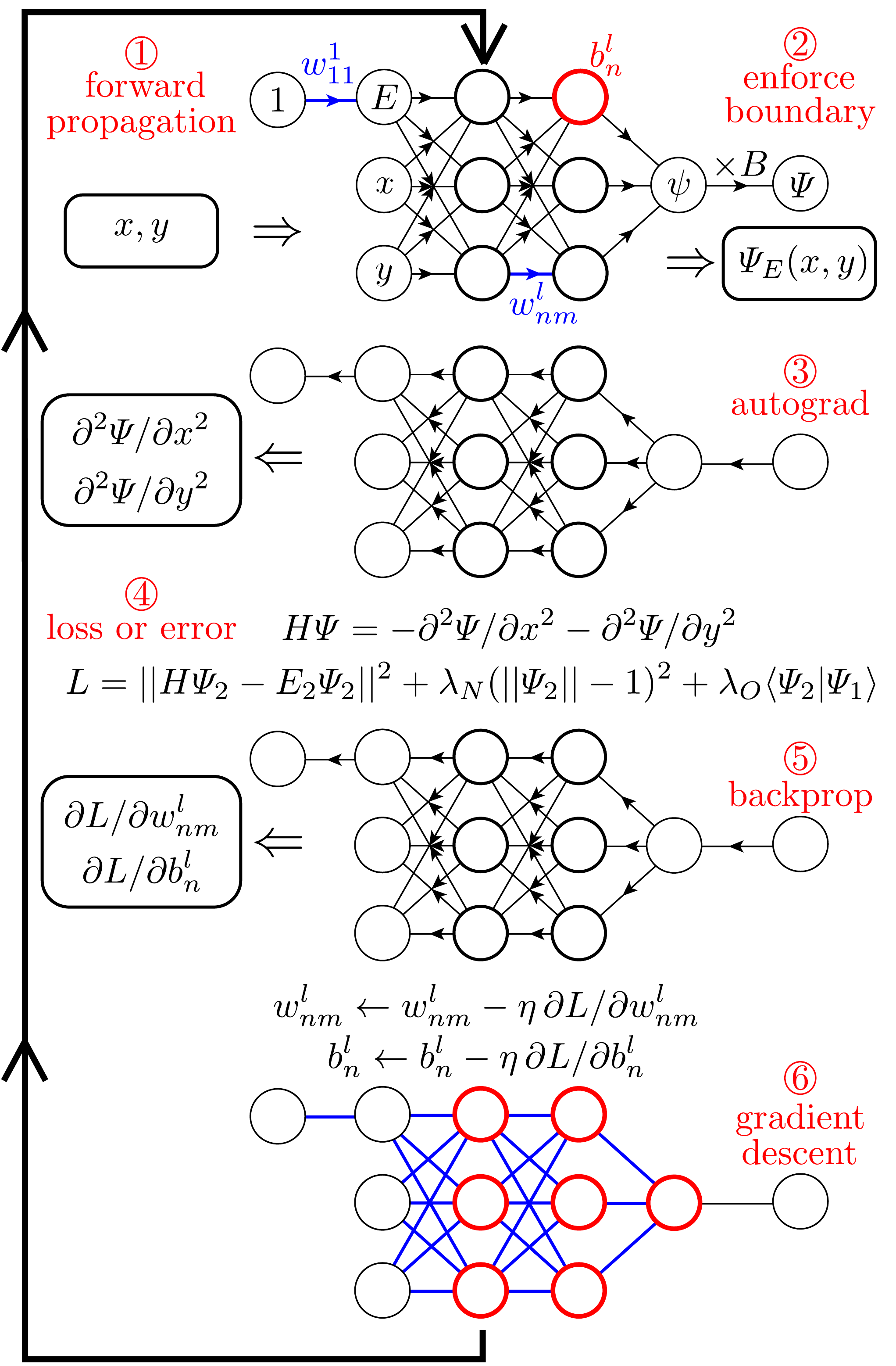}
    \caption{One step of a neural network gradient descent to a second eigenfunction $\varPsi_2$ given a first eigenfunction $\varPsi_1$. Arrows represent weights $w^l_{nm}$ and circles represent biases $b^l_n$. In practice, much of the computation involves reverse-mode automatic differentiation.}
    \label{fig:FlowChart}
\end{figure}

We next extend the one-dimensional JMP algorithm to two dimensions and reformulate it slightly. The network has two hidden layers of neurons with sinusoidal activation functions $\sigma(z) = \sin z$. Inputs are positions $\{x,y\}$ and the constant 1, which is converted into the energy eigenvalue $E = W^1_{11}$ by the adjustable weight $W^1_{11}$, as in Fig.~\ref{fig:FlowChart}. Output is the incomplete eigenfunction $\psi_E (x,y)$, which is multiplied by a function $B(x,y)$ that vanishes on the box's perimeter $\partial \varOmega$ to enforce the boundary conditions and generate the complete eigenfunction 
\begin{equation}
    \varPsi_E(x,y) = B(x,y)\,\psi_E(x,y).
\end{equation}

When finding a second eigenfunction $\varPsi_2$ given a first eigenfunction $\varPsi_1$, the loss
\begin{align}
    L 
    &= L_D + L_N + L_O \nonumber\\
    &= ||H\varPsi_2 - E_2\varPsi_2||^2 \nonumber\\
    &+ \lambda_N (||\varPsi_2|| - 1)^2 \nonumber\\
    &+ \lambda_O \langle \varPsi_2 | \varPsi_1 \rangle \vphantom{(||\varPsi_2|| - 1)^2},
\end{align}
where the Hamiltonian
\begin{equation}
   H = - \frac{\partial^2}{\partial x^2} - \frac{\partial^2}{\partial y^2} + V(x,y),
\end{equation}
and the scalar product
\begin{equation}
   \langle \varPsi | \varPhi \rangle 
   = \int_{-\infty}^\infty \hspace{-1em} dx \int_{-\infty}^\infty \hspace{-1em} dy\, \varPsi(x,y)^* \varPhi (x,y),
\end{equation}
and the norm squared 
\begin{equation}
   || \varPsi ||^2
   = \langle \varPsi | \varPsi \rangle 
   = \int_{-\infty}^\infty \hspace{-1em} dx \int_{-\infty}^\infty \hspace{-1em} dy\, |\varPsi(x,y)|^2.
\end{equation}
Appendix~\ref{NormalizationLoss} discusses different versions of the normalization loss. We take the eigenfunctions to be real, $\varPsi^* = \varPsi$ and $\varPhi^* = \varPhi$, and approximate the integrals as sums, so
\begin{align}
   \langle \varPsi | \varPhi \rangle 
   &\approx  \sum_{m=0}^M \sum_{n=0}^N \delta x\, \delta y\, \varPsi(x_m,y_n)\varPhi(x_m,y_n) \nonumber\\
   &\approx \Delta x\, \Delta y\, \overline{\varPsi_{mn}\varPhi_{mn}}
\end{align}
and
\begin{align}
   || \varPsi ||^2
   &\approx \sum_{m=0}^M \sum_{n=0}^N \delta x\, \delta y\, \varPsi(x_m,y_n)^2 \nonumber\\
   &\approx  \Delta x\, \Delta y\, \overline{\varPsi_{mn}^2}
\end{align}
where $x_m = m\, \delta x = m\, \Delta x / M$ and  $y_n = n\, \delta y = m\, \varDelta y / N$ for the potential well $\varOmega \subset [0,\varDelta x] \times [0, \varDelta y]$, and the overbars indicate  averages.

Minimizing the differential equation loss $L_D$ enforces the Schr\"odinger equation, minimizing the normalization loss $L_N$ discourages trivial zero solutions, and minimizing the orthogonal loss $L_O$ encourages independent solutions. Learning continues until the total loss $L$ and the differential equation loss $L_D$ \textit{and} its rate of change are all small. Appendix~\ref{Implementation} discusses our implementation details.

\begin{figure}[b!]
    \centering
    \includegraphics[width=0.8\linewidth]{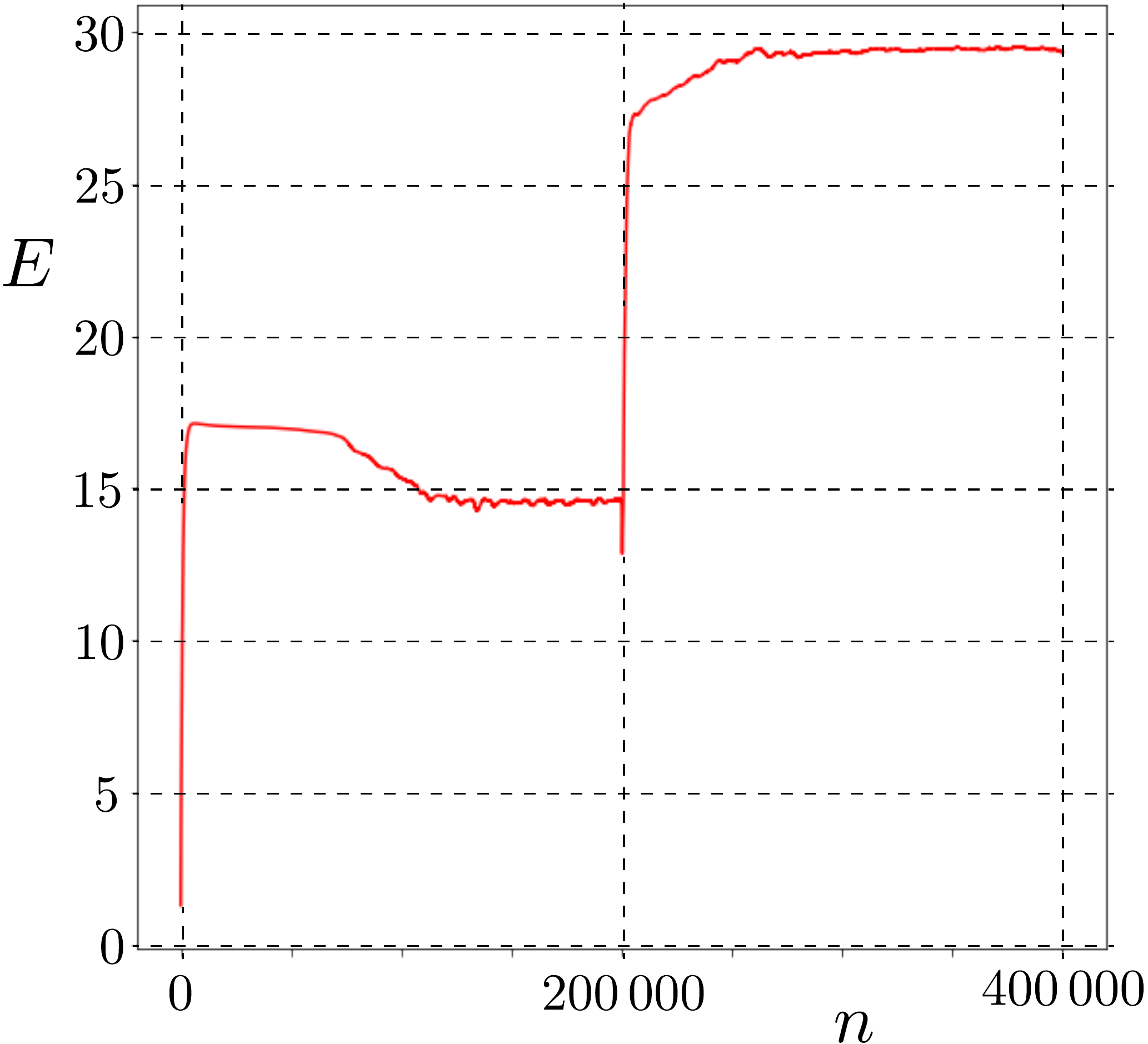}
    \caption{Energy $E$ versus training step $s$ for a rectangle-shaped box.  We determine that the neural network has found an energy eigenvalue by looking for an energy plateau. The jump at the 200\,000th training step is caused by the introduction of the orthogonality loss $L_{O}$, which encourages the neural network to find the next eigenvalue.}
    \label{fig:Ehistory}
\end{figure}

\section{Examples}

We consider potential energy functions of the form
\begin{equation}
    V(x,y) = 
    \begin{cases}
        0, & x,y \in \varOmega, \\
        \infty, & x,y \notin \varOmega,
    \end{cases}
\end{equation}
where $\varOmega$ is the interior of the potential well and $\partial \varOmega$ is its boundary. Our study includes particles trapped in rectangle, elliptic, triangle, and cardioid-shaped potential wells. The rectangle eigenfunctions involve sinusoids, and the elliptic eigenfunctions involve Mathieu functions~\cite{Mathieu1868,Chakraborty2009}, but the triangle and cardioid eigenfunctions are more complicated.

\begin{table*}[bt]
    \caption{\label{tab:example}Examples summary. Boundary functions $B(x,y)$, reference eigenvalues $E^*_n$ computed in Mathematica~\cite{Mathematica}, neural network eigenvalue approximations $E_n$, relative errors $\Delta E_n / E_n$, neural network eigenfunction approximations $\varPsi_n(x,y)$. Red, white, blue palette codes positive, zero, and negative values.} 
    \begin{ruledtabular}
        \renewcommand{\arraystretch}{1.667}
        \begin{tabular}{lccccc}
            boundary function $B$ & reference $E_n$ &  neural net $E_n$ &  error $\Delta E_n / E_n$ & neural net $\varPsi_1$  & neural net $\varPsi_2$   
            \\
            \hline\noalign{\smallskip}
            \centered{  rectangle \\[0.2cm] 
            $\displaystyle  \frac{x}{a}\left(1-\frac{x}{a}\right)\frac{y}{b}\left(1-\frac{y}{b} \right)$ \\[0.2cm] 
            where $a = 1$ and $b = \sqrt{2}$ } &
            \centered{ $14.8$ \\[0.2cm] $29.6$} &
            \centered{ $14.5 \pm 0.1$ \\[0.2cm] $29.4 \pm 0.1$} &
            \centered{ $-2.0\%$ \\[0.2cm] $-0.68\%$ } &
            \centered{ \includegraphics[width=2cm]{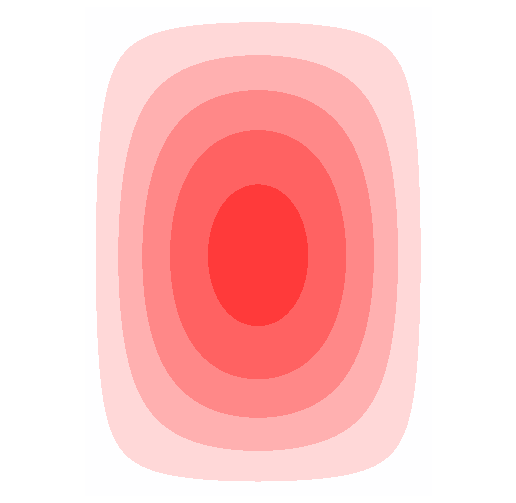} } &
            \centered{ \includegraphics[width=2cm]{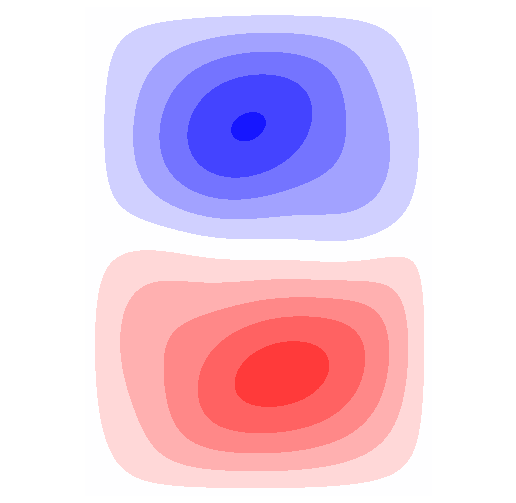}} 
            \\
            \hline\noalign{\smallskip}
            \centered{ ellipse \\[0.2cm] 
            $\displaystyle 1 - \left(\frac{x}{a}\right)^2 - \left(\frac{y}{b}\right)^2$ \\[0.2cm] 
            where $a = 1$ and $b = \sqrt{2}$ } &
            \centered{ $4.32$ \\[0.2cm] $9.13$} &
            \centered{ $4.32 \pm 0.01$ \\[0.2cm] $9.11 \pm 0.08$} &
            \centered{ $\phantom{+}0.00\%$ \\[0.2cm] $-0.22\%$ } &
            \centered{ \includegraphics[width=2cm]{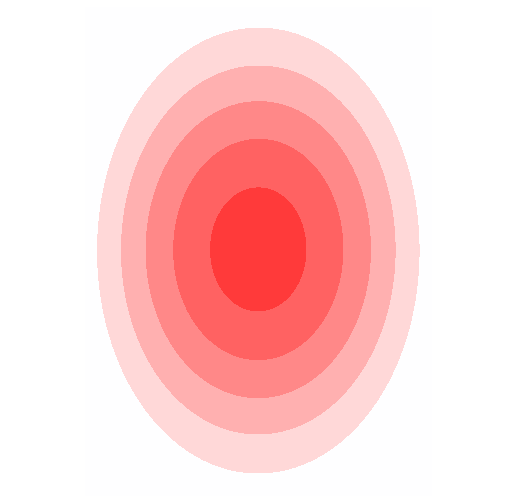} } &
            \centered{ \includegraphics[width=2cm]{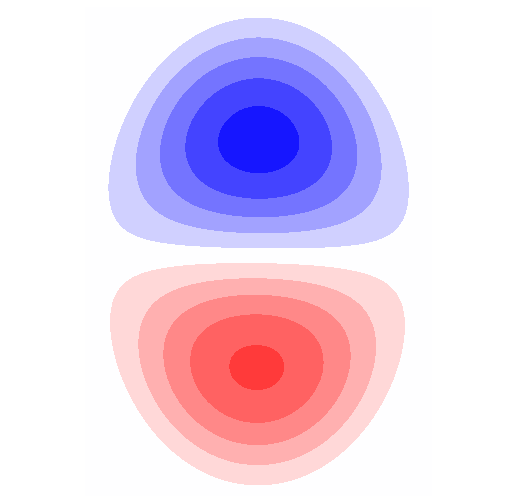} } 
            \\
            \hline\noalign{\smallskip}
            \centered{ triangle \\[0.2cm] 
            $\displaystyle \left(1-\frac{x}{a}\right)\left(\frac{y}{b} - \frac{x}{a} \tan \theta \right)\frac{y}{b} $ \\[0.2cm] 
            where $a = 4$, $b = 4$, and $\theta =  \pi / \sqrt{23}$} &
            \centered{ $4.09$ \\[0.2cm] $8.00$} &
            \centered{ $4.06 \pm 0.01$ \\[0.2cm] $7.98 \pm 0.03$} &
            \centered{ $-0.73\%$ \\[0.2cm] $-0.25\%$}  & 
            \centered{ \includegraphics[width=2cm,trim={4cm 0cm 0cm 7cm},clip]{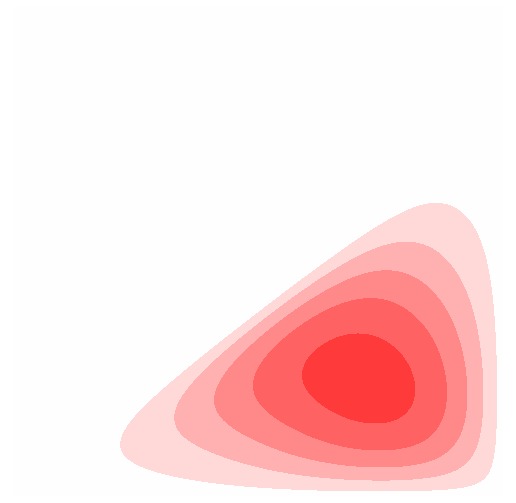} } &
            \centered{ \includegraphics[width=2cm,trim={2.5cm 0cm 0cm 6cm},clip]{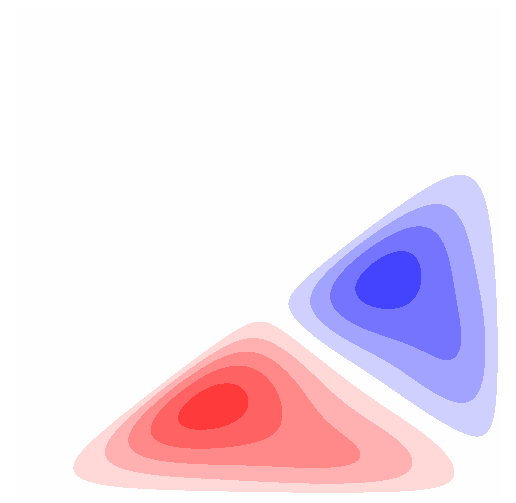} }      \\
            \hline\noalign{\smallskip}
            \centered{ cardioid \color{red}$r = 1 - \delta \sin \theta \Rightarrow$\color{black} \\[0.2cm] 
            $\displaystyle \left(\frac{x}{a}\right)^2+\frac{y}{b}\left(\frac{y}{b}+\delta \right) - \sqrt{\left(\frac{x}{a}\right)^2+\left(\frac{y}{b}\right)^2}$ \\[0.2cm] 
            where $a = 1$, $b = 1$, and $\delta = 1$} &
            \centered{ $4.05$ \\[0.2cm] $9.12$} &
            \centered{ $4.17 \pm 0.15$ \\[0.2cm] $9.13 \pm 0.24$} &
            \centered{ $+3.0\%$ \\[0.2cm] $+0.11\%$ } &
            \centered{ \includegraphics[width=2cm,trim={3.5cm 0.5cm 3.5cm 8.5cm},clip]{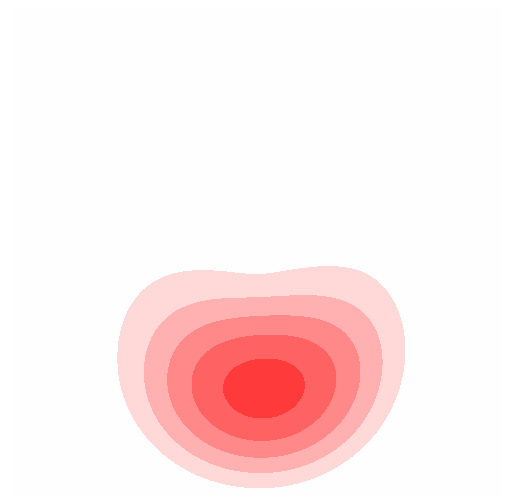} } &
            \centered{ \includegraphics[width=2cm,trim={3.5cm 1cm 3.5cm 8cm},clip]{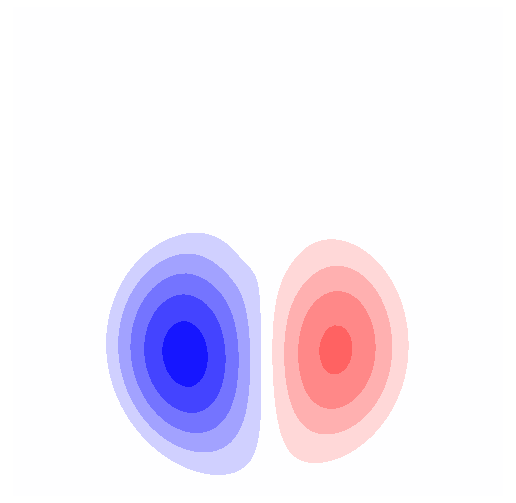} } 
        \end{tabular}
    \end{ruledtabular} 
    \label{table:results}
\end{table*} 

From the classical billiards perspective, rectangle and elliptic-shaped boxes are non-ergodic and integrable, while cardioid-shaped boxes (with polar coordinates boundary $r = 1 - \delta \sin\theta$) have mixed phase spaces when convex ($0\le\delta<1/2$) and are ergodic, mixing, and chaotic when concave ($1/2<\delta\le1$). Triangle-shaped boxes with irrational angles (which are irrational multiples of $\pi$) are ergodic and mixing but not chaotic, but triangle-shaped boxes with one or more rational angles may not even be ergodic~\cite{Lozej2022,Zahradova2022}. (From a spectral analysis perspective, for fixed Dirichlet boundary conditions, the only triangles with explicitly known Laplace spectra are the equilateral $60^\circ-60^\circ-60^\circ$, isosceles right $45^\circ-45^\circ-90^\circ$, and hemi-equilateral $30^\circ-60^\circ-90^\circ$ triangles~\cite{McCartin2008}.) 

We expect the energy eigenvalue spacings of the integrable rectangle and elliptic-shaped boxes to be distributed according to Poisson statistics and the eigenvalue spacings of the chaotic cardioid-shaped boxes to obey Gaussian Orthogonal Ensemble (GOE) statistics~\cite{Casati1980,Bohigas1984}, with the spectral statistics of triangles somewhere in between~\cite{Lozej2022}.

\section{Results}

We initiate our study by constructing a fully-connected feed-forward neural network with $1+2$ inputs, 2 hidden layers containing 150 to 200 neurons each (depending on the potential's complexity), and 1 output. To train the neural network, we generate 100 random $\{x,y\}$ points within the box and feed them into the network according to the algorithm outlined in Fig.~\ref{fig:FlowChart}. We repeat for about $10^5$ training loops or epochs.

As in Fig.~\ref{fig:Ehistory}, the neural network adjusts its weights and biases to converge to the ground-state energy and remains at that energy plateau until the orthogonality term $L_{O}$ is added to the loss function. Once it is present, the neural network leaves the ground-state energy plateau to find the energy associated with the first-excited-state.

Table~\ref{table:results} summarizes the results. The reference energies were numerically computed in Mathematica~\cite{Mathematica} (and checked exactly for the rectangle). The energy eigenvalue uncertainty is the wiggle of the energy plateau as it rings down to its mean value, and the relative error $\Delta E / E$ is the estimated energy minus the reference energy divided by the reference energy. The eigenfunction color palette stretches from fully saturated red (for $\varPsi > 0)$ to fully saturated blue (for $\varPsi < 0$) via completely unsaturated white (for $\varPsi = 0$). 

 Smaller boxes have higher energy states, as momentum $p \propto 1 / \lambda$ implies energy $E = p^2 / 2m \propto 1 / \lambda^2$, where $\lambda$ is the quantized wavelength. Thus, the Table~\ref{table:results} scaling parameters $a,b$ can be adjusted to keep the energy eigenvalues in a convenient numerical range.

The neural network implementing the two-dimensional JMP algorithm successfully approximates the ground and first-excited energy eigenvalues and eigenfunctions for particles confined to a wide range of boxes without assuming or imposing the boxes' symmetries. Higher-order excited states can be obtained similarly by adding further orthogonality terms to the loss function. Additional training can increase accuracy. Most difficult and most impressive is the cardioid, or pinched-circle-shaped box, whose second and third states are nearly degenerate, with the pinch at top breaking the degeneracy and making the eigenfunction with the horizontal node slightly more energetic than the eigenfunction with the vertical node.

\section{Other Applications}

Related applications include acoustic and electromagnetic cavities and Laplacian spectra. The time-independent Schr\"odinger Eq.~\ref{TISE} can be written
\begin{equation}
    (-\nabla^2 + V) \varPsi = E\varPsi.
\end{equation}
Inside the hard-walled infinite wells, $V = 0$ and
\begin{equation}
    \nabla^2 \varPsi = -E\varPsi.
\end{equation}
This is the same as the Helmholtz equation
\begin{equation}
    \nabla^2 f = -k^2 f,
\end{equation}
which can describe waves on membranes with clamped edges in two dimensions~\cite{Rayleigh1945} and electromagnetic waves in conducting cavities in three dimensions~\cite{Stockmann1990, Nockel1997}. A special case is the Laplace equation
\begin{equation}
    \nabla^2 f = 0,
\end{equation}
which is central to the mathematical problem of Laplacian eigenvalues for planar domains with Dirichlet boundary conditions~\cite{McCartin2008}.

\section{Conclusions}

We have demonstrated that the JMP algorithm, when extended to two dimensions, enables neural networks to solve the time-independent Schr\"odinger equation and find quantum energy eigenvalues and eigenfunctions for both classically regular and irregular billiards systems. Such capability is yet another example of physics-informed machine learning navigating dynamical systems that exhibit both order and chaos~\cite{Choudhary2020}. 

This success is proof-of-concept that a simple feed-forward neural network, incorporating physics intuition in its loss function, can solve complicated eigenvalue problems, even if well-established state-of-the-art numerical methods are currently faster or more accurate. Two-dimensional JMP neural networks have much potential. Future work includes exploiting spatial symmetries to reduce the number of training points and generalizing to continuous and three dimensional potential wells.

\begin{acknowledgments}
This research was supported by the Office of Naval Research grant N00014-16-1-3066 and a gift from United Therapeutics Corporation.
\end{acknowledgments}

\appendix

\section{Gradient Descent} \label{GradientDescent}

\subsection{Dynamical Analogue}
Gradient descent of a neural network weight $w$ is like a point particle of mass $m$ at position $x$ sliding with viscosity $\gamma$ on a potential energy surface $V(x)$. Newton's laws imply
\begin{equation}
    m \ddot x  = F_x = - \frac{dV}{dx} - \gamma \dot x,
\end{equation}
where the overdots indicate time differentiation. For large viscosity $|m \ddot x| \ll|\gamma \dot x | $ and
\begin{equation}
    \gamma \dot x \sim -\frac{dV}{dx},
\end{equation}
so the velocity
\begin{equation}
    \dot x \sim - \frac{1}{\gamma} \frac{dV}{dx}.
\end{equation}
Position evolves like the Euler update
\begin{equation}
    x \leftarrow x + dx = x + \dot x\, dt \sim x - \frac{1}{\gamma} \frac{dV}{dx} dt.
\end{equation}
With position $x = w$, height $V = L$, and learning rate $\eta = dt/\gamma$, a neural network weight evolves like
\begin{equation} \label{momFree}
    w \leftarrow w - \eta \frac{dL}{dw},
\end{equation}
and similarly for a bias. Increasing the learning rate makes the loss surface more ``slippery", while decreasing the learning rate makes the loss surface more ``sticky", and variable learning rates may expedite gradient descent to a global minimum. Model \textit{stochastic} gradient descent by buffeting the sliding particle with noise.

\subsection{Newton's Method}
Alternately, seek minima of the loss function $L(w)$ by seeking roots of its derivative $L^\prime(w)$ according to the Newton-Raphson method of extending the tangent to the intercept and stepping
\begin{equation}
    w \leftarrow w - \frac{L^\prime(w)}{L^{\prime\prime}(w)}.
\end{equation}
If a minimum at $w_m$ is approximately quadratic, so
\begin{equation}
    L(w) = \frac{1}{2}a (w - w_m)^2 + c,
\end{equation}
then nearby Newton's method reduces to 
\begin{equation}
    w \leftarrow w - \frac{1}{a} L^\prime(w)
    = w - \eta \frac{dL}{dw},
\end{equation}
where the learning rate $\eta = 1 / a = 1 / L^{\prime\prime}(w)$ is inverse to the curvature.
\section{Normalization Loss}\label{NormalizationLoss}

For the normalization loss, Jin et al.~\cite{Jin2022C} propose
\begin{align}
    L_N 
    &= (\langle \varPhi | \varPhi \rangle - M / \Delta x)^2 \nonumber\\
    &= c\, (|| \varPsi ||^2 - 1)^2,
\end{align}
where $c = (M/\Delta x)^2$ and $|\varPsi\rangle = |\varPhi\rangle / \sqrt{M / \Delta x}$. However, an alternative loss is
\begin{equation}
    L_N = c\, (|| \varPsi || - 1)^2.
\end{equation}
The latter is arguably simpler, while the related function
\begin{equation}
    L_N = c\, (|| \varPsi ||^2 - 1)
\end{equation}
is problematic because it can be positive or negative.
\newpage

\section{Implementation Details}\label{Implementation}

\begin{figure}[tb]
    \centering
    \includegraphics[width=1.0\linewidth]{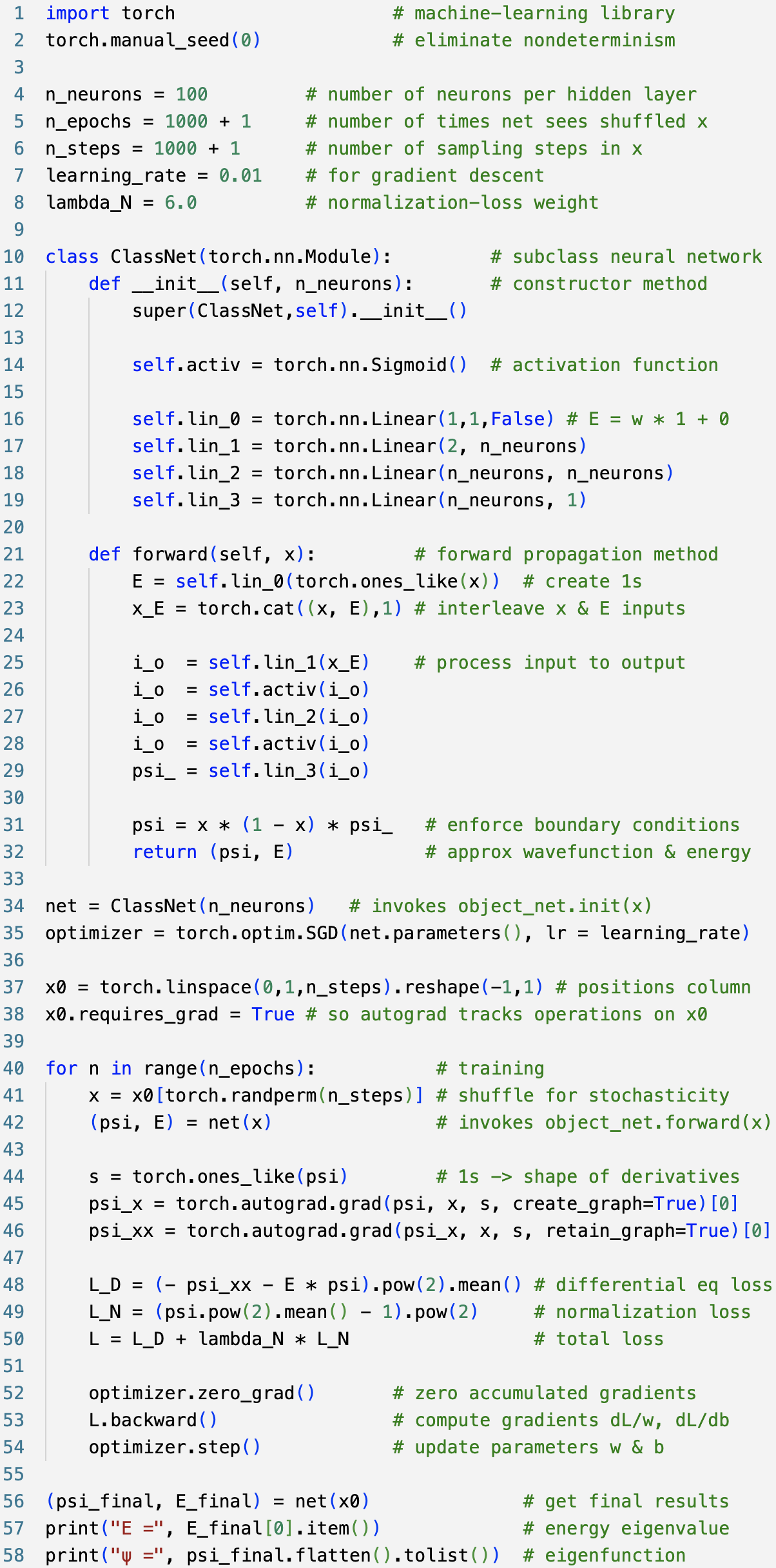}
    \caption{Example Python code of a neural network learning to model a particle in a one-dimensional box $\varOmega = [0,1]$. Returns ground state energy within about $1\%$ of the exact value.}
    \label{fig:PyTorchCode}
\end{figure}

The Fig.~\ref{fig:PyTorchCode} Python sample code implements a simple neural network with sigmoid activation functions that learns the ground state energy eigenvalue and eigenfunction of a particle in a one-dimensional box $\varOmega = [0,1]$.  Our PyTorch machine-learning library implementation uses tensors (multidimensional rectangular arrays of numbers) throughout. \texttt{ClassNet} (lines 10-32) defines the network architecture (lines 14-19), implements a forward pass (lines 21-32), and enforces the Dirichlet boundary conditions (line 31). After instantiating an object of the class \texttt{object\underline{ }net} and initializing the stochastic gradient descent optimizer (lines 34-35),  variable \texttt{x} is a list or columnar array of equally-spaced positions inside the box with \texttt{autograd} tracking its operations (lines 37-38).  

The for-loop (lines 40-54) manages the neural network training, first shuffling the \texttt{x} values (line 41) and then asking \texttt{object\underline{ }net} for the latest eigenfunction and energy eigenvalue approximations (line 42). The \texttt{grad} function invokes \texttt{autograd} to compute the Laplacian (lines 44-46). The \texttt{.pow()} and \texttt{.mean()} methods help compute the loss function (lines 48-50). 

The \texttt{.backward()} method also invokes \texttt{autograd} and computes the gradients of the loss with respect to weights and biases, which are then updated by the optimizer (lines 52-54). More generally, the \texttt{.backward()} method computes the \texttt{.grad} attribute of all tensors that have \texttt{requires\underline{ }grad = True} in the computational graph of which \texttt{loss} is the final leaf and the inputs are the roots; then \texttt{optimizer} iterates through the list of weight and bias parameters it received when initialized and, wherever a tensor has \texttt{requires\underline{ }grad = True}, it subtracts the value of its gradient (multiplied by the learning rate) stored in its \texttt{.grad} property. 

Final energy eigenvalue and eigenfunction are extracted as numbers and printed (lines 56-58). This working example returns a ground state energy within about $1\%$ of the exact value.
    
\newpage

\bibliographystyle{unsrt}

\providecommand{\noopsort}[1]{}\providecommand{\singleletter}[1]{#1}%

\end{document}